# COMPARISON OF SECURE AND HIGH CAPACITY COLOR IMAGE STEGANOGRAPHY TECHNIQUES IN RGB AND YCBCR DOMAINS

Hemalatha S[1], U Dinesh Acharya[2], Renuka A[3]

[1,2,3]Department of Computer Science and Engineering, Manipal Institute of Technology, Manipal University, Manipal, Karnataka, India
[1]hema.shama@manipal.edu;[2]dinesh.acharya@manipal.edu;
[3]renuka.prabhu@manipal.edu

## ABSTRACT

*Steganography is one of the methods used for secret communication. Steganography attempts to hide the existence of the information. The object used to hide the secret information is called as cover object. Images are the most popular cover objects used for steganography. Different techniques have to be used for color image steganography and grey scale image steganography since they are stored in different ways. Color image are normally stored with 24 bit depth and grey scale images are stored with 8 bit depth. Color images can hold large amount of secret information since they have three color components. Different color spaces namely RGB (Red Green Blue), HSV (Hue, Saturation, Value), YUV, YIQ, YCbCr (Luminance, Chrominance) etc. are used to represent color images. Color image steganography can be done in any color space domain. In this paper color image steganography in RGB and YCbCr domain are compared. The secret information considered is grey scale image. Since RGB is the common method of representation, hiding secret information in this format is not secure.*

## KEYWORDS

*Steganography, Integer Wavelet Transform (IWT), MSE, PSNR, RGB, YCbCr, Luminance, Chrominance*

## 1. INTRODUCTION

Information security is essential in the field of secure communication. Steganography is used for secure information exchange. Communication between two parties is always subject to unauthorized interception. Steganography is one of the methods used to send secret information such that there is no evidence that the opponent can intercept and try decoding the information. In steganography some medium such as image, audio or video is used to hide secret information. The signal used to hide secret information is called as cover object. Stego-object is the cover along with the hidden information [1].

The most popular cover objects used for steganography are images. In image steganography color image steganography finds more importance than grey scale image steganography because color images have large space for information hiding. Different color spaces namely RGB (Red Green Blue), HSV (Hue, Saturation, Value), YUV, YIQ, YCbCr (Luminance, Chrominance) etc. are used to represent color images. Color image steganography can be done in any color space domain. One of the best representations for steganography is YCbCr. Human eye is very sensitive to changes in luminance but not in chrominance. So small changes in chrominance part cannot alter the overall image quality much [2,3]. Luminance component is Y and Cb, Cr are the blue and red chrominance components respectively. These are shown in Figure 1.

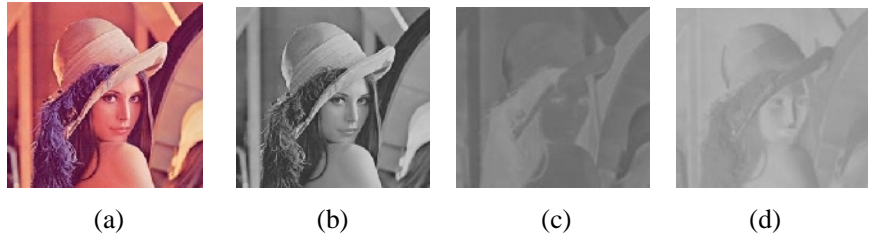

    (a)           (b)           (c)           (d)

Figure 1 (a) Lena, (b) Y component (luminance) of (a), (c) Cb component (chrominance blue) of (a), (d) Cr component (chrominance red) of (a).

The conversion formulae are used to convert values from one color space to another color space [3]. The conversion to YCbCr from RGB is as follows:

$$Y = (77/256)R + (150/256)G + (29/256)B \qquad (1.1)$$

$$Cb = -(44/256)R - (87/256)G + (131/256)B + 128 \qquad (1.2)$$

$$Cr = (131/256)R - (110/256)G - (21/256)B + 128 \qquad (1.3)$$

The opposite transformation is

$$R = Y + 1.371(Cr - 128) \qquad (1.4)$$

$$G = Y - 0.698(Cr - 128) - 0.336(Cb - 128) \qquad (1.5)$$

$$B = Y + 1.732(Cb - 128) \qquad (1.6)$$

In this paper color image steganography in RGB and YCbCr domain are compared. Image steganography may be in spatial domain or transform domain. Spatial domain techniques embed secret information in the pixel values directly. Here the advantage is the simplicity. But they suffer ability to bear signal processing operations like filtering, compression, noise insertion, cropping etc. In transform domain methods, the cover image is transformed into different domain and then the transformed coefficients are modified to hide secret information. The stego image is obtained by transforming back the modified coefficients into spatial domain. High ability to face signal processing operations is the advantage of transform domain methods [1], but they are computationally complex. However in steganography since robustness and security are the two main characteristics, transform domain techniques are more preferable. In this paper Integer Wavelet Transform (IWT) is used to hide secret images in the color cover image.

The stego signal is highly robust if it can withstand manipulations such as filtering, cropping, rotation, compression etc. Security is the inability of an attacker to detect the information even if the existence of the information is realized [4]. If the stego image is closer to the cover image, then it is very secure. It is measured in terms of Peak Signal to Noise Ratio (PSNR).

$$PSNR = 10 \log \frac{L^2}{\sqrt{MSE}} dB \qquad (1.7)$$

where L = peak value, MSE = Mean Square Error.

$$MSE = \frac{1}{N} \sum_{i=1}^{N} |X_i - X_i'|^2 \qquad (1.8)$$

where X = original value, X' = stego value and N = number of samples.

The larger the PSNR value the higher the security since it indicates minimum difference between the original and stego values. So no one can detect the hidden information.

The Wavelet Transform gives time-frequency representation of the signal. IWT is an efficient approach to lossless compression. The coefficients in this transform are represented by finite precision numbers which allows for lossless encoding. It maps integers to integers. But in Discrete Wavelet Transform, the output coefficients are not integers when the input consists of integers, as in the case of images. Thus the perfect reconstruction of the original image is not possible. Since IWT maps integers to integers the output can be completely characterized with integers and so perfect reconstruction is possible. In case of IWT, the LL sub-band is a close copy with smaller scale of the original image while in the case of DWT the LL sub-band is slightly distorted. This is shown in Figure 2 [5].

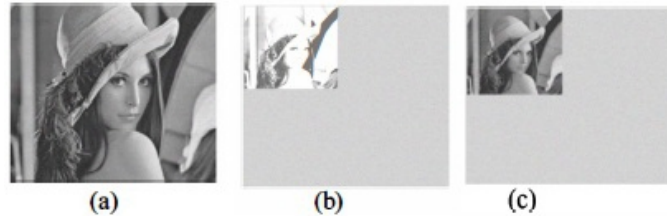

Figure 2 (a) Original image Lena. (b) LL sub band in one level DWT (c) LL sub-band in one level IWT.

## 2. RELATED WORK

In this section some of the recent colour image steganography techniques are reviewed.
A new approach based on LSB is proposed by Masud, Karim S.M., Rahman, M.S., Hossain, M.I. [6]. The hidden information is encrypted by a secret key and then it is stored into different position of LSB of image. This provides very good security. Xie, Qing., Xie, Jianquan., Xiao, Yunhua [7]., proposed a method to hide the information in all RGB planes based on HVS (Human Visual System). The quality of the stego image is degraded.

Vector Quantization (VQ) table is used by Sachdeva S and Kumar A,[8], to hide the secret message which enhances the capacity and also stego size. For hiding text messages Roy, S., Parekh, R., [9] proposed an improved steganography approach. They hide the text message within lossless RGB images. This fails to face signal processing operations. Mandal, J.K., Sengupta, M. proposed minimum deviation of fidelity based data embedding technique [10]. Here the replacement positions are chosen randomly between LSB and up to fourth bit towards MSB to modify two bits per byte. Rubab, S., Younus, M. [11], embedded text message in color image using DWT. Blowfish encryption technique is used to encrypt the text message. This is a complex method. Nabin Ghoshal et al., (SSCIA) [12] used DFT to hide image.

In the proposed work the actual secret images are not hidden in the cover image. Just by hiding the keys the secret images are transmitted. YCbCr technique is implemented in our previous works [13] and [14]. In [13], one grey scale is hidden in color image using Discrete Wavelet Transform (DWT) and IWT. In [14], the capacity is improved by hiding two grey scale images in one color image using IWT only. This improves the quality of the stego image. In the proposed method, different encryption method is used to encrypt the keys so that technique becomes more secure.

## 3. PROPOSED METHOD

In the proposed method, the cover is 256x256 color image. Two grey scale images of size 128 x128 are used as secret images. The following sections describe the embedding and extracting methods for steganography in RGB and YCbCr domains.

### 3.1. Steganography in RGB Domain

The color cover image is considered in the normal RGB representation. The three components red green and blue are extracted. Two grey scale images are hidden in two color components. The experiment is conducted first by hiding in green and blue components and then by hiding in red and green components. In the second case the quality of the stego image is slightly better than that in the first case but the quality of the extracted secret images are slightly poor. This comparison is done by calculating the PSNR values. When red and blue components are used for hiding secret images the results are same as those with red and green components. So the results are shown when the secret images are hidden in green and blue components. The following sub sections explain the embedding and extracting processes in the RGB domain.

### 3.1.1. Embedding Process

- Let the cover image be C and let two secret images be S1 and S2.
- Single level IWT of secret-images S1, S2 and green and blue components of C, are obtained.
- Four sub-bands corresponding to low and high frequency coefficients are obtained. They are termed as LL, LH, HL and HH sub bands.
- LL sub band of green component is used to hide one secret image and LL sub band of blue component is used to hide another secret image as follows:
    - The sub-images CgLL and S1LL of green component of C and S1 are divided into non-overlapping blocks of size 2x2. i.e., $BCg_k$ =(1 ≤ k < nc) and $BS1_i$ =(1 ≤ i < ns), where $BCg_k$, $BS1_i$ represent $k^{th}$ block in CgLL, $i^{th}$ block in S1LL respectively and nc is the total number of 2x2 blocks in CgLL and ns is the total number of 2x2 blocks in S1LL.
    - Every block $BS1_i$, is compared with block $BCg_k$. The pair of blocks which have the least Root Mean Square Error is determined. The address of the best matched block $BCg_k$ for the block $BS1_i$ is stored in a key. Let it be K1.
    - Similarly, the key K2 is obtained by considering the LL sub band of blue component of C and LL sub band of S2.
    - The two keys are then encrypted and then hidden in the cover image. The high frequency components LH, HL and HH of green component of C are used to hide the key K1 and K2 is hidden in high frequency components of blue component of C. The steps are as follows:
        - The binary image is constructed using the middle bit planes of the higher frequency components of the transformed image of green component of C.
        - The Key K1 is compressed.
        - The bits of the compressed key K1 are hidden in the middle bit planes of the higher frequency components of the transformed image of green component of C. (fourth and higher bit planes are used in the experimentation.)
        - The inverse IWT of the resulting image is obtained to get the green component of the stego image.
    - Similarly K2 is hidden in the middle bit planes of the higher frequency components of the transformed image of blue component of C.
    - Then inverse IWT is applied to get the blue component of the stego image.

### 3.1.2. Extracting Process

- Decompose the stego image into red, green and blue components.

- Find the integer wavelet transform of green and blue components of the stego image separately.
- The binary image is obtained using the higher frequency components of the green and blue components of the transformed image.
- The encrypted, compressed keys K1 and K2 are present in the middle bit planes of the higher frequency components of green and blue components of the transformed image respectively.
- Decrypt and decompress them to obtain the original keys K1 and K2.

Then secret images are obtained with the help of keys K1 and K2 as follows:

- Divide LL sub band of green component into non overlapping blocks of size 2x2.
- The blocks that are the nearest approximation to the original blocks of S1LL are obtained using K1.
- Rearrange the blocks to obtain S1LLnew.
- Obtain the secret image S1 by considering S1LHnew, S1HLnew and S1HHnew as zero matrices of dimension S1LLnew and then by obtaining inverse IWT of S1new.
- Similarly obtain S2 from LL sub band of blue component using K2.

### 3.2. Steganography in YCbCr Domain

The normal color image in RGB is transformed into YCbCr color space. The following sub sections explain the embedding and extracting processes in the YCbCr domain.

### 3.2.1. Embedding Process

- Let the cover image be C and let two secret images be S1 and S2.
- The cover image C is represented in YCbCr color space.
- Single level IWT of secret-images S1, S2 and Cb, Cr component of C, are obtained.
- Four sub-bands corresponding to low and high frequency coefficients are obtained. They are termed as LL, LH, HL and HH sub bands.
- One secret image is hidden in LL sub band of Cb and another secret image is hidden in LL sub band of Cr. The steps are as follows:
    - The sub-images CbLL and S1LL of Cb component and S1 are divided into non-overlapping blocks of size 2x2. i.e., $BCb_k$ =(1 ≤ k < nc) and $BS1_i$ =(1 ≤ i < ns), where $BCb_k$, $BS1_i$ represent $k^{th}$ block in CbLL, $i^{th}$ block in S1LL respectively and nc is the total number of 2x2 blocks in CbLL and ns is the total number of 2x2 blocks in S1LL.
    - Every block $BS1_i$, is compared with block $BCb_k$. The pair of blocks which have the least Root Mean Square Error is determined. The address of the best matched block $BCb_k$ for the block $BS1_i$ is stored in a key. Let it be K1.
    - Similarly, the key K2 is obtained by considering the LL sub band of Cr and LL sub band of S2.
    - The two keys are then encrypted and then hidden in the cover image. The high frequency components LH, HL and HH of Cb are used to hide the key K1 and K2 is hidden in high frequency components of Cr. The steps are as follows:
        - The binary image using the middle bit planes of the higher frequency components of the transformed image of Cb is constructed.
        - The Key K1 is compressed.
        - The bits of the compressed key K1 are hidden in the middle bit planes of the higher frequency components of the transformed image of Cb. (fourth and higher bit planes are used in the experimentation.)

- ▪ The inverse IWT of the resulting image is obtained to get the Cb component of the stego image.
  - o Similarly K2 is hidden in the middle bit planes of the higher frequency components of the transformed image of Cr.
  - o The resultant image is represented in RGB color space to obtain stego image G.

### 3.2.2. Extracting Process

The steps to extract secret images from the Cb and Cr components of the stego image are as follows:

- The stego image G is represented in YCbCr color space. Let it be GyGcbGcr
- IWT of Gcb and Gcr components is obtained.
- The binary image is constructed using the higher frequency components of the transformed image separately.
- The middle bit planes of the higher frequency components of Gcb and Gcr contain the encrypted, compressed keys K1 and K2 respectively.
- Decrypt and decompress them to obtain the original keys K1 and K2.

Then secret images are obtained with the help of keys K1 and K2 as follows:

- GcbLL is divided into non overlapping blocks of size 2x2.
- The blocks that are the nearest approximation to the original blocks of S1LL are obtained using K1.
- Rearrange the blocks to obtain S1LLnew.
- Obtain the secret image S1 by considering S1LHnew, S1HLnew and S1HHnew as zero matrices of dimension S1LLnew and then by obtaining inverse IWT of S1new.
- Similarly obtain S2 from GcrLL using K2.

## 4. EXPERIMENTAL RESULTS

The algorithm is tested in MATLAB. The wavelet tool box is used. The lifting wave *cdf 2.2* is used to find the integer wavelet transform. Various test images are used to test the algorithm. Figure 3 shows original cover and secret images. The cover images used are "baboon" and "peppers" (Figure 3(a) and 3(b)), each of size 256X256. The secret images considered are "earth" and "football" (Figure 3(c) and 3(d)), each of size 128X128.

The results of steganography in RGB domain are shown in Figure 4. Figure 4(a) and 4(b) show the stego images. Figure 4(c) and 4(d) show the extracted secret images from "baboon". Figure 4(e) and 4(f) show the extracted secret images from "peppers".

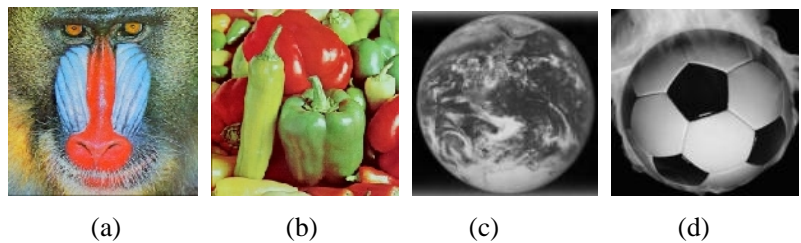

(a)  (b)  (c)  (d)

Figure 3 Cover and secret images: (a) cover (baboon)  (b) cover (peppers)  (c) secret (earth)  (d) secret (football)

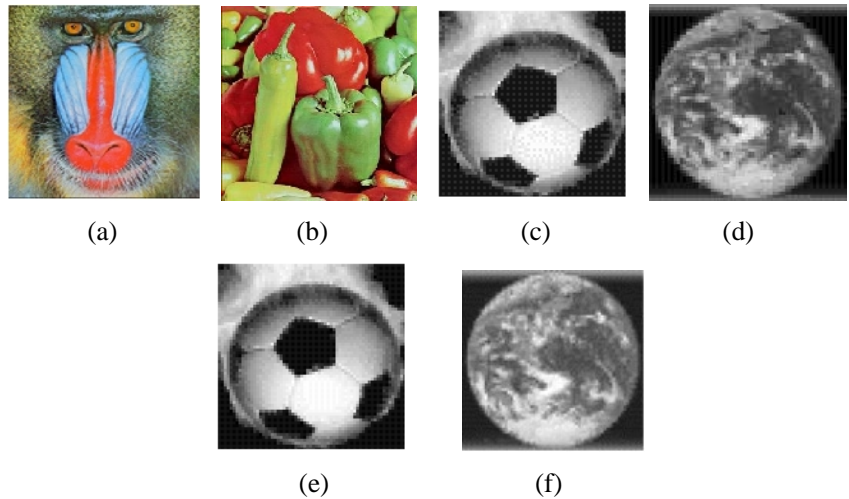

Figure 4 Stego and extracted secret images from RGB steganography: (a) stego (baboon), (b) stego (peppers), (c)-(f) extracted secret images: (c) football from baboon   (d) earth from baboon (e) football from peppers   (f) earth from peppers

The results of steganography in YCbCr domain are shown in Figure 5. Figure 5(a) and 5(b) show the stego images. Figure 5(c) and 5(d) show the extracted secret images from "baboon".  Figure 5(e) and 5(f) show the extracted secret images from "peppers".

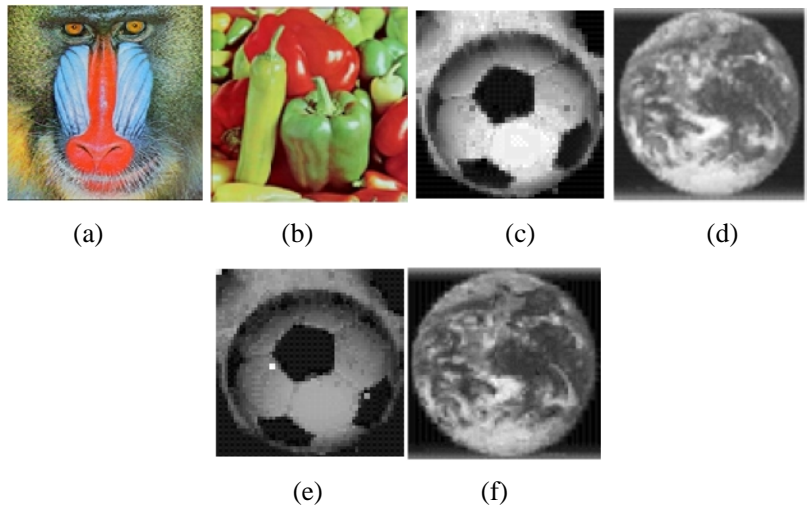

Figure 5 Stego and extracted secret images from YCbCr steganography: (a) stego (baboon), (b) stego (peppers), (c)-(f) extracted secret images: (c) football from baboon   (d) earth from baboon (e) football from peppers   (f) earth from peppers

Table 1 shows the PSNR value of the stego images in the two steganographic methods. Table 2 shows the PSNR values of the extracted secret images in the two steganographic methods.

Table 1. PSNR (in dB) of the stego images

| COVER IMAGE | STEGANOGRAPHIC METHOD | |
|---|---|---|
| | RGB | YCbCr |
| peppers | 47 | 41 |
| baboon | 48 | 41 |

Table 2. PSNR (in dB) of the extracted secret image

| SECRET IMAGE | STEGANOGRAPHIC METHOD | |
|---|---|---|
| | RGB | YCbCr |
| football | 27 | 32 |
| earth | 28 | 33 |

## 5. CONCLUSIONS

In this paper, two color image steganography methods are compared along with the experimental results. Since in color images large amount of information can be hidden color image steganography finds importance in the information security field. The results show that the quality of the stego images are good in RGB domain by comparing the PSNR values. But RGB method may be detected more easily than YCbCr method since RGB is the normal standard representation for color images. And RGB steganography method requires more execution time than YCbCr method. The PSNR values show that the quality of extracted secret images are good in YCbCr method. But only PSNR value is not the quality measuring metric as the visual appearance of the resultant images reveal. Some other parameters must be used to mesure the security or qlity of the steganography technique. This will be the factor for our future work.